\documentclass[dvips]{article}
\usepackage{icrctc07}

\title{Search for Coincidences in Time and
  Arrival Direction of Auger Data with Astrophysical Transients}
\shorttitle{Search for coincidences between Auger data and astrophysical
  transients}

\authors{Luis 
Anchordoqui$^1$, for the Pierre Auger Collaboration$^2$}
\shortauthors{Pierre Auger Collaboration}
\afiliations{$^1$
Department of Physics,
University of Wisconsin-Milwaukee, P.O. Box 413, Milwaukee, WI 53201, USA\\
$^2$Pierre Auger Observatory, Av. San Mart\'{\i}n Norte 304, (5613) Malarg\"ue, Argentina}
\email{anchordo@uwm.edu}

\abstract{The data collected by the Pierre Auger Observatory are analyzed to
  search for coincidences between the arrival directions of
  high-energy cosmic rays and the positions in the sky of
  astrophysical transients. Special attention is directed towards
  gamma ray observations recorded by NASA's Swift mission, which have
  an angular resolution similar to that of the Auger surface
  detectors. In particular, we check our data for evidence of a signal
  associated with the giant flare that came from the soft gamma repeater 
  1806-20 on December 27, 2004.}

\begin{document}
\maketitle
\section{Introduction}

It has long been known that two of the highest energy cosmic rays came
from directions that are within the error boxes of two remarkable
gamma ray bursts (GRBs) detected by BATSE, with a delay of roughly 10
months after the bursts~\cite{Milgrom:1995um}.  However, all searches
for coincidences between the arrival direction of ultra-high energy
cosmic rays (UHECRs) and GRBs from the Third BATSE catalogue yield a
negative result~\cite{Stanev:1996qc}. These searches, of course, may
have been distorted by the poor angular resolution (about 3 degrees)
of the GRB measurements. A sensitive anisotropy analysis is now
feasible using data collected by the Swift
mission~\cite{Gehrels:2004am} and the Pierre Auger
Observatory~\cite{Abraham:2004dt}. In the first part of this paper we
present the results of such an analysis.

In the second part of the paper, we search for a signal from the
direction of the soft gamma repeater (SGR) 1806-20, which on December
27, 2004 emitted a hyperflare that saturated many satellite
$\gamma$-ray detectors~\cite{Mereghetti:2005dt}. This unique flare
lasted about 5 minutes (the duration of the initial spike was $\sim
0.2~{\rm s}$), had a peak luminosity of $\sim 2 \times 10^{47}~{\rm
  erg/s},$ and a total energy emission of $\sim 5 \times
10^{46}$~erg~\cite{Woods:2006nc}. The model proposed is that of a
``magnetar'' (i.e., a neutron star with a huge magnetic field,
$\vec{B} \sim 10^{15}~{\rm G}$) located on the far side of our Galaxy
(at a distance $\approx 15$~kpc~\cite{Corbel}).  The origin of the
flare can be explained as global crustal fractures due to
$\vec{B}$-field rearrangements liberating a high flux of $X$-rays and
$\gamma$-rays~\cite{Thompson:1995gw}.  The exceptional energetics of
this hyperflare makes of SGR 1806-20 an attractive candidate source of
UHECRs, high energy neutrinos, and gravitational
waves~\cite{Ioka:2005er}. Given that the source is in our Galaxy, if
it were to generate high energy neutrons a significant fraction of
them could arrive at Earth before decaying and would point back to the
source.  Moreover, neutrons that decay in flight would produce
antineutrinos that inherit directionality~\cite{Anchordoqui:2003vc}.
Searches for neutrino and gravitational wave emission have been
reported by the AMANDA~\cite{Achterberg:2006az},
AURIGA~\cite{Baggio:2005xi}, and LIGO Scientific~\cite{Abbott:2007px}
collaborations. In all these searches the data revealed no significant
signals.

\section{Cosmic Ray Data Sample}

In our analysis we use data collected with the Pierre Auger
Observatory, located in Argentina at Southern latitude $ 35.2^\circ$
and Western longitude $69.3^\circ$. We consider events with zenith
angle $\theta < \theta_{\rm max} = 60^{\circ}$, detected from January
1, 2004 to April 1, 2007.  We employ the reconstruction procedure
discussed elsewhere~\cite{Bonifazi:2005ns}.  A total of 609,161
CRs have passed the selection criteria.

\begin{figure*}
\begin{center}
\includegraphics [width=1.\textwidth]{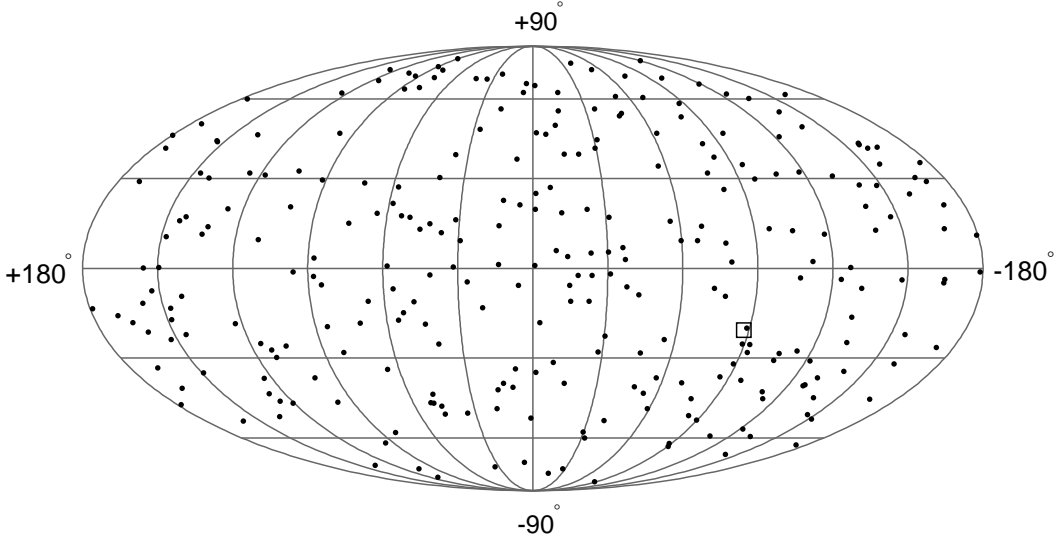}
\end{center}
\caption{The dots indicate the positions of 284 well-localized
  GRBs (equatorial coordinates J2000) observed from January 1, 2004,
  to April 1, 2007. 192 of them are within the f.o.v. of Auger with a
  zenith angle $< 60^{\circ}$, and 234 with a zenith angle
  $<80^{\circ}$. The boundary of Auger f.o.v. for $\theta_{\rm
    max}=60^{\circ}$ is defined by declination = $24.8^\circ.$
 The
  position of SGR 1806-20 is indicated by an open square.}
\label{GRB-sky}
\end{figure*}

\section{UHECRs and GRBs}

A catalogue of 284 GRBs observed with an accuracy of better than
1$^{\circ}$ (from Januray 1, 2004 to April 1, 2007) was compiled using
data primarily from the Swift mission complemented by measurements
from additional GRB observing satellites, including HETE, INTEGRAL,
and others. Out of the total GRB sample, 192 bursts are within the
field of view (f.o.v.) of Auger and only 62
were in the Auger f.o.v. at the time of their bursts, i.e.
$\theta_{\rm GRB}<\theta_{\rm max}.$ The GRB sky distribution is given
in Fig.~\ref{GRB-sky}. As expected~\cite{Meegan:1992xg}, they cover
the whole sky isotropically.

\begin{figure}
\begin{center}
\includegraphics [width=0.5\textwidth]{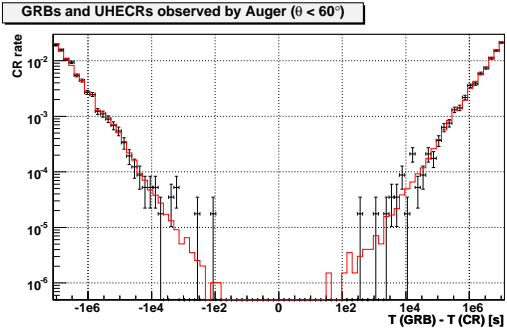}
\end{center}
\caption{Rates of CR events as a function of the
  difference between the GRB time and the CR arrival time. Data
  falling within $30^\circ$ of a GRB are indicated by the histogram
  and within $5^\circ$ by the points. For clarity, statistical errors
  are only shown for the $5^{\circ}$-distribution.}
\label{GRB-time}
\end{figure}

We bin CR events coming from directions defined by spherical caps of
radii $\vartheta_{\rm sep} = 5^\circ$ and $\vartheta_{\rm sep} =
30^\circ$ around each GRB position. We determine the total number of
coincidence candidates by counting the number of CRs found within each
of the specified cones.  The number of coincidence candidates for
$30^\circ$ is ${\cal N}_{30^\circ} = 1,980,577$ whereas for $5^\circ$,
${\cal N}_{5^\circ} = 57,053$.  Note that for $30^\circ$ the number of
coincidence candidates is larger than the total number of CRs in the
sample, since a given CR may lie within $30^\circ$ of more than one
GRB. Differences between the observation times of the GRBs and the
arrival times of CR events in the same angular bin were determined. In
Fig.~\ref{GRB-time}, we show the rates of CR events as function of the
GRB-CR time difference for the two angular radii $\vartheta_{\rm sep}$
around the positions of GRBs.  We consider a 100-day period before and
after the GRB observation. The histogram corresponds to a $30^{\circ}$
cut and the points corresponds to a $5^{\circ}$ cone. No significant
excess after the time of the bursts is evident in the data.

\begin{figure}
\begin{center}
\includegraphics [width=0.5\textwidth]{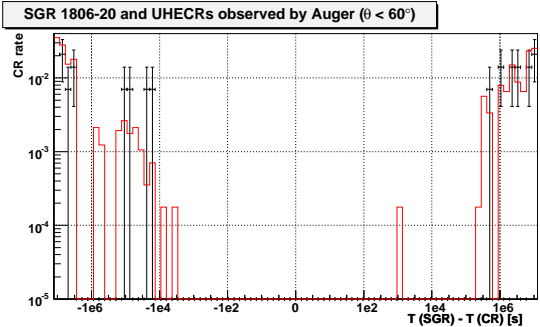}
\end{center}
\caption{Rates of CR events from the direction of SGR 1806-20 as a
  function of the $\gamma$-CR time difference. Conventions are the
  same as in Fig.~\ref{GRB-time}.}\label{SGR-time}
\end{figure}

\section{UHECRs from SGR 1806-20?}

The giant flare of December 27, 2004 from SGR 1806-20 represents one
of the most intriguing events captured in almost three decades of
monitoring the $\gamma$-ray sky. Such an energetic event clearly
constitutes a potential candidate for acceleration of UHECRs.
Secondary neutrons can be produced in collisions of relativistic
protons (and nuclei) with the ambient plasma. Interestingly, those
produced with an energy $E > 10^{18}~{\rm eV}$ have a boosted
$c\tau_n$ sufficiently large to serve as Galactic messengers.  The
decay mean free path of a neutron is $c \Gamma_n \overline \tau_n =
9.15 (E/10^{18}~{\rm eV})~{\rm kpc}$, the lifetime being boosted
from its rest frame value, $\overline \tau_n = 886~{\rm s},$ to its lab
value by $\Gamma_n = E/m_n.$ Because of the exponential depletion, 
about 20\% of the
neutrons survive the trip at $10^{18}~{\rm eV}$, and about 58\% at
$10^{18.5}~{\rm eV}$.

The location of the source, right ascension $18{\rm h}\ 08{\rm m}\
39.34{\rm s}$ and declination $-20^{\circ}\ 24'\ 39.7''$, is within
the f.o.v. of Auger, and below $60^\circ$ for about 9 hours per day. 
At the flaring emission 21:30:26.5 UTC, its
zenith angle was $\theta_{\rm SGR} = 70.3^\circ,$ and it remained for
a 50 minute interval above the horizon. Unfortunately, this is outside
the currently best understood region of the detector, i.e. $\theta_{\rm
  max} = 60^\circ$.  

We have repeated the analysis described in the previous section for
this exceptional burst.  In Fig.~\ref{SGR-time} we show the results of
such an analysis, indicating that no significant excess in the CR flux
is evident after the burst.  (The number of events coming from the
direction of SGR 1806-20 within a $30^\circ$ cone is ${\cal
  N}_{30^\circ} = 5,596$ whereas for $5^\circ$, ${\cal N}_{5^\circ} =
139$.)

\begin{figure}
\begin{center}
\includegraphics [width=0.49\textwidth]{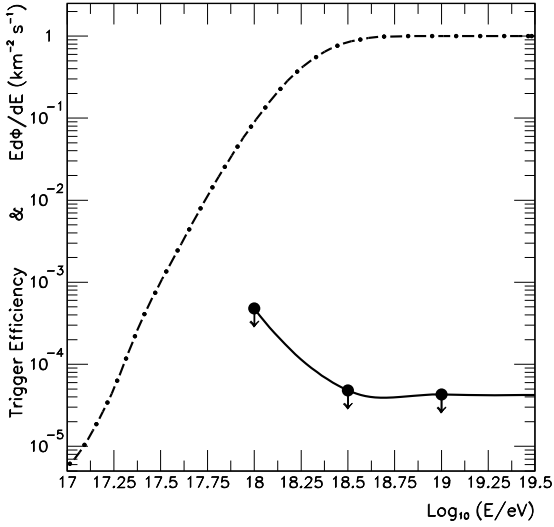}
\end{center}
\caption{The rising dot-dashed line indicates the trigger efficiency
  of Auger for $\theta = 70^\circ.$ The arrowed circles indicate upper
  limits on the energy weighted flux of neutrons from SGR 1806-20
  (valid in a logarithmic interval $\Delta =1$).}
\label{limit}
\end{figure}

By extending our data analysis to higher zenith angles, we have
verified that no events have been observed within a $5^\circ$ cone
during the $T = 300$~s of the flare, where $\theta_{\rm SGR} \approx
70^\circ.$ The abscence of a signal can then be exploited to place an
upper bound on the primary neutron flux, without assumptions on the
Galactic magnetic field. To do so, we must determine the effective
detection area $A$ and the trigger efficiency $\epsilon (E)$. We adopt
the elementary hexagonal cell approach discussed in
Ref.~\cite{Allard:2005he}.  On December 27, 2004 an average of 364
Auger hexagones were fully active, each counting for $1.95~{\rm km}^2$
on the ground (which amounts to $0.66~{\rm km}^2$ as seen with an
angle of $\approx 70^\circ$).  Hence, the experiment presented $A
\simeq 239~{\rm km}^2$ to the potential "beam" of neutrons emanating
from the source. The trigger efficiency for $70^\circ$ is shown in
Fig.~\ref{limit}.

Using the diffuse flux of cosmic rays we estimate a background $\ll
1$, thus Poisson statistics implies an upper bound of 3.09 events at
95\% CL from neutron fluxes, $d\Phi/dE$~\cite{Feldman:1997qc}.
Equivalently, for some interval $\Delta,$
\begin{equation}
T\,\, A\,\,\int_\Delta  \,\,dE\,\, \frac{d\Phi}{dE} \, 
\epsilon (E)  < 3.09 \,\,.
\end{equation}
In a logarithmic interval $\Delta$ where a
single power law approximation (for the integrand) is valid, we
obtain~\cite{Anchordoqui:2002vb}
\begin{equation}
\left. E_0 \,\, \frac{d\Phi}{dE}\right|_{E_0}  < \frac{3.09}{T\,\,A\,\, 
\epsilon(E_0)} \, \,\,,
\end{equation}
where $E_0$ is the energy at the center of the logarithmic interval
and we have taken $\Delta = 1$ (corresponding to one $e$-folding of
energy). The 95\%CL upper limits on the energy weighted flux of
neutrons are shown in Fig.~\ref{limit}.

\section{Conclusion}

We used the Auger data sample to search for UHECRs which are
consistent with the position and time of astrophysical transients. No
such coincidences were found in the data.

\end{document}